\documentclass[onecolumn]{emulateapj}
\usepackage{graphicx}
\usepackage{amsmath, amssymb}
\slugcomment{}
\shortauthors{Suto et al.}
\shorttitle{Validity of HSE }

\newcommand{\gap}{\ \ \ }
\newcommand{\del}{\partial}

\newcommand{\bs}{\boldsymbol}
\newcommand{\pd}[2]{\frac{\partial#1}{\partial#2}}
\newcommand{\dt}[1]{\frac{\partial #1}{\partial t}}
\newcommand{\dr}[1]{\frac{\partial #1}{\partial r}}
\newcommand{\dth}[1]{\frac{\partial #1}{\partial\theta}}
\newcommand{\dph}[1]{\frac{\partial #1}{\partial\varphi}}
\newcommand{\Mtot}{M_{\rm tot}}
\newcommand{\Mth}{M_{\rm therm}}
\newcommand{\Mrot}{M_{\rm rot}}
\newcommand{\Mstr}{M_{\rm stream}}
\newcommand{\Macc}{M_{\rm accel}}
\newcommand{\Mrand}{M_{\rm rand}}
\newcommand{\Maniso}{M_{\rm aniso}}

\newcommand{\Mcross}{M_{\rm cross}}
\newcommand{\vt}{v_\theta}
\newcommand{\vp}{v_\varphi}
\newcommand{\sigmarr}{\sigma_{rr}}
\newcommand{\sigmart}{\sigma_{r\theta}}
\newcommand{\sigmarp}{\sigma_{r\varphi}}

\newcommand{\sigmatt}{\sigma_{\theta\theta}}
\newcommand{\sigmatp}{\sigma_{\theta\varphi}}

\newcommand{\sigmapp}{\sigma_{\varphi\varphi}}
\newcommand{\rhog}{\rho_{\mathrm{gas}}}
\newcommand{\rhod}{\rho_{\mathrm{dm}}}
\newcommand{\rhos}{\rho_{\mathrm{star}}}
\newcommand{\rhot}{\rho_{\mathrm{tot}}}

\begin{document}
%
\title{ Validity of Hydrostatic Equilibrium in Galaxy Clusters from Cosmological Hydrodynamical Simulations }

\author{
 Daichi Suto\altaffilmark{1}, 
 Hajime Kawahara\altaffilmark{2}, 
 Tetsu Kitayama\altaffilmark{3}, 
 Shin Sasaki\altaffilmark{2},\\
 Yasushi Suto\altaffilmark{1,4,5}
 and Renyue Cen\altaffilmark{5} }
\email{daichi@utap.phys.s.u-tokyo.ac.jp}

\altaffiltext{1}{Department of Physics, The University of Tokyo, 
Tokyo 113-0033, Japan}
\altaffiltext{2}{Department of Physics, Tokyo Metropolitan University,
  Hachioji, Tokyo 192-0397, Japan}
\altaffiltext{3}{Department of Physics, Toho University,  Funabashi,
  Chiba 274-8510, Japan}
\altaffiltext{4}{Research Center for the Early Universe, School of Science, The
University of Tokyo, Tokyo 113-0033, Japan}
\altaffiltext{5}{Princeton University Observatory, Princeton, NJ 08544}

\begin{abstract}
We examine the validity of the hydrostatic equilibrium (HSE) assumption
for galaxy clusters using one of the highest-resolution cosmological
hydrodynamical simulations. We define and evaluate several {\it
effective} mass terms corresponding to the Euler equations of the gas
dynamics, and quantify the degree of the validity of HSE in terms of the
mass estimate.  We find that the mass estimated under the HSE assumption
(the HSE mass) deviates from the true mass by up to $\sim 30$ \%. This
level of departure from HSE is consistent with the previous claims, but
our physical interpretation is rather different. We demonstrate that the
inertial term in the Euler equations makes a negligible contribution to
the total mass, and the overall gravity of the cluster is balanced by
the thermal gas pressure gradient and the gas acceleration term.  Indeed
the deviation from the HSE mass is well explained by the acceleration
term at almost all radii. We also clarify the confusion of previous work
due to the inappropriate application of the Jeans equations in
considering the validity of HSE from the gas dynamics extracted from
cosmological hydrodynamical simulations.

\end{abstract}

\section{Introduction}

Clusters of galaxies are important sources of various cosmological and
astrophysical information especially on the formation history of the
large scale structure and the estimates of cosmological parameters
\citep[][for a recent review]{allen11}. Among others, mass of clusters
is one of the most fundamental quantities in virtually all studies. The
most conventional method is based on X-ray observations of the
intracluster medium (ICM) combined with the assumption that the ICM is
in hydrostatic equilibrium (HSE) with the total gravity of the cluster
(We call the mass estimated by this method ``the HSE mass''). It is
unlikely, however, that the HSE assumption strictly holds, especially
for unrelaxed clusters given their on-going dynamical
evolution. Therefore it is important to examine the validity of the HSE
assumption, which has been mostly assumed just for simplicity.  The
quantitative analysis of its validity and limitation is directly related
to the applicability to the future scientific opportunities, including
upcoming X-ray missions such as extended R\"ontgen Survey with Imaging
Telescope Array\footnote{http://www.mpe.mpg.de/eROSITA} (eROSITA) and
ASTRO-H\footnote{http://astro-h.isas.jaxa.jp}, and observations of the
Sunyaev-Zel'dovich effect performed by Atacama Cosmology
Telescope\footnote{http://www.princeton.edu/act} (ACT) and South Pole
Telescope\footnote{http://pole.uchicago.edu} (SPT).

The validity of the HSE assumption for observed clusters may be examined in a straightforward fashion by comparison of the HSE mass with the cluster mass estimated by other methods.  In this respect, gravitational lensing is particularly suited because it directly probes the total gravitational mass without any assumption on the dynamical state of dark matter. On the other hand, the lensing observations require a high angular resolution of the background galaxy images and are feasible only for a limited number of clusters located at $z\lesssim0.5$. In addition, the estimated lensing mass correspond to the cylindrical mass along the line of sight, and may include an extra contribution not associated with the cluster itself. Previous studies \citep[e.g.,][for recent ones]{mahdavi08, zhang08} show that the HSE mass is smaller by approximately 20 percent on average than the lensing mass, suggesting either HSE or lensing, or even both, should be systematically biased.

Another method to examine the validity of the HSE assumption that we pursue in this paper is to use numerical simulations, which enable us to make a detailed and critical comparison of the simulated data against the model prediction. Therefore we can locate the origin of systematic bias, if any, of the HSE assumption. This is useful because we may be able to apply the correction to the observational data eventually.

Of course there are a number of previous studies of HSE using simulated clusters, but their results do not seem to be converged.  For instance, let us focus on a couple of recent papers \citep{fang09,lau09} that studied systematic errors in the HSE mass using the {\it same} set of 16 clusters simulated by \citet{nagai07}. \citet{fang09} analyzed the gas particle data on the basis of the Euler equations, and evaluated the {\it effective} mass terms corresponding to several different terms in the equations.  \citet{lau09} performed basically the same analysis, but used the Jeans equations instead of the Euler equations despite the fact that they considered the gas particles in the simulated clusters.  Both reached the similar conclusion that the HSE mass underestimates the true mass of clusters systematically by $\sim$ 10--20 \%. Nevertheless their physical interpretations of the origin of the bias are very different; \citet{fang09} claimed that the coherent rotation of gas plays a significant role as an additional support against the gravity, while \citet{lau09} concluded that the random gas motion is responsible for the departure from HSE, and the gas rotation makes a relatively negligible contribution.

The purpose of the present paper is to clarify the theoretical method to examine HSE of the simulated clusters, and then to revisit its validity using a high-resolution hydrodynamical simulation by \citet{cen12}. In particular, we compare the two different analysis formulations adopted by \citet{fang09} and \citet{lau09}, and argue that the Euler equations, rather than the Jeans equations modified by a gas pressure gradient term \citep{rasia04,lau09}, should be used in analyzing the gas dynamics.

The rest of the paper is organized as follows. Section 2 formulates our method to evaluate the validity of the HSE assumption. We pay a particular emphasis on the comparison between the Euler and Jeans equations in analyzing the gas dynamics. The derivation of those sets of equations starting from the Boltzmann equations is summarized in Appendix A. Section 3 briefly describes the simulated cluster, and then presents our analysis results of the validity of HSE. Finally our conclusion is summarized in Section 4. The analysis based on the Jeans equations using dark matter particles is shown in Appendix B.

\section{Theoretical Formulation}

\subsection{Method to Examine the Validity of HSE based on the Euler Equations}

Our analysis method to discuss the validity of HSE using gas in a simulated cluster is based on the Euler equations:
\begin{equation}
\dt{\bs{v}}+(\bs{v}\cdot\nabla)\bs{v}=-\frac{1}{\rhog}\nabla{p}-\nabla{\phi},
\label{euler}
\end{equation}
where $\phi$ is the gravitational potential, and $\rhog$, $\bs{v}$ and
$p$ are the density, velocity and pressure of gas.  While the Jeans
equations do not describe the gas dynamics in their original form,
\citet{rasia04} and \citet{lau09} added the gas pressure gradient term
to the Jeans equations, and adopted the resulting equations in analyzing
simulated clusters. We argue in the next subsection that this is not
justified, and present the result based on the Jeans equations but using
collisionless dark matter particles in Appendix B, for reference.

We define the total mass $\Mtot$ of a cluster inside a volume $V$ as
\begin{equation}
\Mtot=\int_V d^3x\ \rhot,
\end{equation}
where $\rhot$ is the total density of the cluster. For the simulated
cluster considered throughout this paper, $\rhot$ consists of densities
of gas, dark matter and stars, i.e., $\rhot=\rhog+\rhod+\rhos$. The
total mass can be rewritten in terms of $p$ and $\bs{v}$ using Poisson's
equation and Gauss's theorem:
\begin{equation}
\Mtot=\frac{1}{4\pi G}\int_{\del V} d\bs{S}\cdot\nabla\phi
=\frac{1}{4\pi G}\int_{\del V} d\bs{S}\cdot
\left[-\frac{1}{\rhog}\nabla p -(\bs{v}\cdot\nabla)\bs{v}
-\dt{\bs{v}}\right],
\label{gravE}
\end{equation}
where $\del V$ is the surface surrounding the volume $V$ and we have
used equation (\ref{euler}) in the second equality. Now the total mass
is evaluated by the gas quantities alone, without any knowledge on dark
matter and stars. This is why the present method is applicable, in
principle, to the X-ray data of galaxy clusters.

If we adopt a spherical surface as $\del V$, the total mass can be
decomposed into the following four {\it effective} mass terms:
\begin{equation}
\Mtot=\Mth+\Mrot+\Mstr+\Macc,
\label{eulerdecomp}
\end{equation}
where
\begin{equation}
\Mth=-\frac{1}{4\pi G}\int_{\del V} dS\ \frac{1}{\rhog}\dr{p},
\label{mthE}
\end{equation}
\begin{equation}
\Mrot=\frac{1}{4\pi G}\int_{\del V} dS\ \frac{v_\theta^2+v_\varphi^2}{r},
\label{mrotE}
\end{equation}
\begin{equation}
\Mstr=-\frac{1}{4\pi G}\int_{\del V} dS\ \left[v_r\dr{v_r}+\frac{v_\theta}{r}\dth{v_r}+\frac{v_\varphi}{r\sin\theta}\dph{v_r}\right],
\label{mstrE}
\end{equation}
and 
\begin{equation}
\Macc=-\frac{1}{4\pi G}\int_{\del V}  d S\ \dt{v_r}.\label{maccE}
\end{equation}
We emphasize here that the above set of equations does not assume
spherical symmetry of the system; we just take a spherical surface as
the integral surface and write down equation (\ref{gravE}) in spherical
coordinates.

The first term, $\Mth$, originates from the thermal pressure gradient of
gas. If the gas motion is negligible, $\Mth$ should be equal to the
total mass $\Mtot$, and thus regarded as a cluster mass estimated under
the HSE assumption. In other words, the difference between $\Mtot$ and
$\Mth$ is a quantitative measure of the departure from the HSE
assumption.

The inertial term $(\bs{v}\cdot\nabla)\bs{v}$ in the Euler equations
reduces to $\Mrot$ and $\Mstr$. If there exists a coherent rotational
motion around the center of the cluster, $\Mrot$ can be interpreted as
the centrifugal force term. Without such a motion, however, the local
tangential velocity of different directions at different locations on
the sphere could make $\Mrot$ significantly large.  During the course of
the cluster evolution, gas generally falls toward the center of the
cluster with larger streaming speed with increasing radius from the
center. In this case, $\Mstr$ becomes {\it negative} and cannot be
neglected.  On the other hand, it becomes positive and/or negligible,
especially in the innermost region where the gas velocity is more
randomized than that in the outer regions.

Finally the acceleration term, $\Macc$, corresponds to the $- \partial
v_r/\partial t$ term in the Euler equations, and becomes
positive/negative when gas is decelerating/accelerating.

All the mass terms are invariant with respect to the choice of the axis
of the spherical coordinates, but are not necessarily positive.  Note
also that $\Mrot$ and $\Mstr$, corresponding to the inertial term, are
not invariant with respect to the Galilean transformation. Thus we
evaluate those in the center-of-mass frame of the entire simulated
cluster.

\subsection{Comparison with Analysis Methods Adopted by Previous Work}

The set of basic equations that we adopt in this paper is essentially
identical to that of \citet{fang09}, except the fact that they
interpreted the difference between $\Mtot$ and $\Mth+\Mrot+\Mstr$ as
``turbulent gas motion'' while we call it the acceleration term $\Macc$
as directly implied from equation (\ref{euler}), and evaluate it from
the residual, $\Macc=\Mtot-\Mth-\Mrot-\Mstr$.

We are not sure why they ascribed the term to the turbulent motion.  It
is true that part of the gas acceleration would be due to the turbulent
gas motion, but not entirely. Furthermore the numerical simulation does
not include any physical processes directly related to the turbulent
motion. Even if the turbulent motion might be important for real
clusters, it should come from some physics below the subgrid scales that
cannot be properly resolved in the numerical simulation.  Effects of gas
random motion above the resolved scales should be included in $\Mrot$
and $\Mstr$.  We note, however, that the different interpretation of
$\Macc$ does not affect at all the conclusion of \citet{fang09} that gas
rotation is the most important term to describe the origin of departure
from HSE in their simulated clusters. As we will show below, this is not
consistent with our result.

In previous literature, the Jeans equations are sometimes used in
analyzing the gas motion in simulated clusters. For that purpose, a
thermal pressure gradient term is added by hand to the basic equations
\citep[e.g.,][]{rasia04,lau09}.  Assuming the steady state, i.e.,
$\del\bs{v}/\del t=0$, the Jeans equations in $r$-direction
(\ref{jeansR}) is now replaced by the following equation:
\begin{eqnarray}
\begin{split}
\left[v_r\dr{}+\frac{\vt}{r}\dth{}+\frac{\vp}{r\sin\theta}\dph{}\right]v_r
+\frac{1}{\rhog}\left[\frac{1}{r}\dth{(\rhog\sigmart^2)}
+\frac{1}{r\sin\theta}\dph{(\rhog\sigmarp^2)}\right]
+\frac{\sigmart^2\cot\theta}{r}\\
\ \ \ \ \ \ -\frac{\vt^2+\vp^2}{r}=-\frac{1}{\rhog}\dr{(\rhog\sigmarr^2)}
-\frac{2\sigmarr^2-\sigmatt^2-\sigmapp^2}{r}-\frac{1}{\rhog}\dr{p}-\dr{\phi},
\end{split}
\label{laujeans}
\end{eqnarray}
where $\sigma_{ij}^2$ denotes the $ij$-component of the velocity dispersion tensor. 

For example, \citet{lau09} converted each term of equation (\ref{laujeans}) to the corresponding mass term; see their equations (6) to (11). Then they interpreted that the first two terms on the right hand side of the above equation
originate from ``random gas motion''.  We suspect that the implicit assumption underlying equation (\ref{laujeans}) is that the ICM consists of two distinct gas components; the thermal gas and the unthermalized one such as the cold gas accreting along with galaxies into clusters. While it may be reasonable to take into account both components separately when dealing with {\it real galaxy clusters}, it
remains to be justified if the extra pressure gradient term can be added by hand to the conventional Jeans equations assuming that the two components share the same density $\rhog$ (In Appendix A we show that diagonal components of the velocity dispersion tensor in the Jeans equations correspond to thermal pressure in the Euler equations). It should also be recalled that the current numerical simulations are
performed entirely on the basis of the Euler equations.  Therefore, for the sake of consistency, it is appropriate to use the Euler equations, instead of equation (\ref{laujeans}), in evaluating the mass terms of the \textit{simulated galaxy clusters}. The proper treatment of the unthermalized gas in numerical simulations is certainly an important issue, but is beyond the scope of the present paper. 
\section{Application to a Simulated Cluster}

We apply the analysis method in the previous section and quantitatively examine the validity of the HSE assumption. For that purpose, we use a simulated cluster of \citet{cen12}. It is simulated with an Eulerian adaptive mesh refinement (AMR) code, Enzo \citep{bryan99, bryannorman99, oshea04, joung09}. Although our current analysis exploits three-dimensional profiles of the cluster which cannot be observed in reality, we are interested here in the validity of HSE itself. The application to the X-ray observation will be discussed separately elsewhere.

\subsection{Hydrodynamical Simulation}

We briefly describe the main features of the current simulation, and refer readers to \citet{cen12} for more detail. The simulation was run first with a low resolution mode in a periodic box of 120 $h^{-1}$ Mpc on a side. Then a region centered on a cluster with a mass of $\sim 2 \times 10^{14} h^{-1} M_\odot$ was resimulated with a higher resolution in an adaptively refined manner. The size of the refined region is $21 \times 24 \times 20 (h^{-1} \rm{Mpc})^3$. The mean interparticle separation and the dark matter particle mass in the refined region are 117 $h^{-1}$ kpc comoving and $1.07 \times 10^8 h^{-1} M_\odot$, respectively.

Star particles are created according to the prescription of
\citet{cen92}. Their typical mass is $\sim 10^6 M_\odot$. The simulation
includes a metagalactic UV background \citep{haardt96}, shielding of UV
radiation by neutral hydrogen \citep{cen05} and metallicity-dependent
radiative cooling \citep{cen95}. While supernova feedback is modeled
following \citet{cen05}, AGN feedback is not included in this
simulation. The cosmological parameters used in this simulation are
$(\Omega_b, \Omega_m, \Omega_\Lambda, h, n_s, \sigma_8) =$ (0.046, 0.28,
0.72, 0.70, 0.96, 0.82), following the WMAP7-normalized $\Lambda$CDM
model \citep{komatsu11}.

Then a cluster is identified and the cubic box of a side of $3.8 h^{-1}$
Mpc surrounding the entire cluster is extracted from the simulation
data. The dark matter and stars are represented by particles, and the
temperature and density of gas are given on the 520$^3$ grids (the grid
length is $7.324 h^{-1}{\rm kpc}$).

The radius $r_{500}$ of the cluster is $\sim 640 h^{-1}$ kpc ($r_{500}$
is defined so that the mean density inside $r_{500}$ is 500 times the
critical density of the universe).  The center-of-mass velocity of the
cluster within $r_{500}$ is set to vanish.  The total mass $M_{500}$
within $r_{500}$ is $\sim 2 \times 10^{14} M_\odot$. The average ICM
temperature at $r_{500}$ is $\sim$ 2 keV, and the circular speed there
is $v_{500}=\sqrt{GM_{500}/r_{500}}\sim 1000$ km s$^{-1}$.

\begin{figure}
\begin{center}
\includegraphics[width=7.2cm]{fig1a.eps}\hspace{1cm}
\mbox{\raisebox{0cm}{\includegraphics[width=5.6cm]{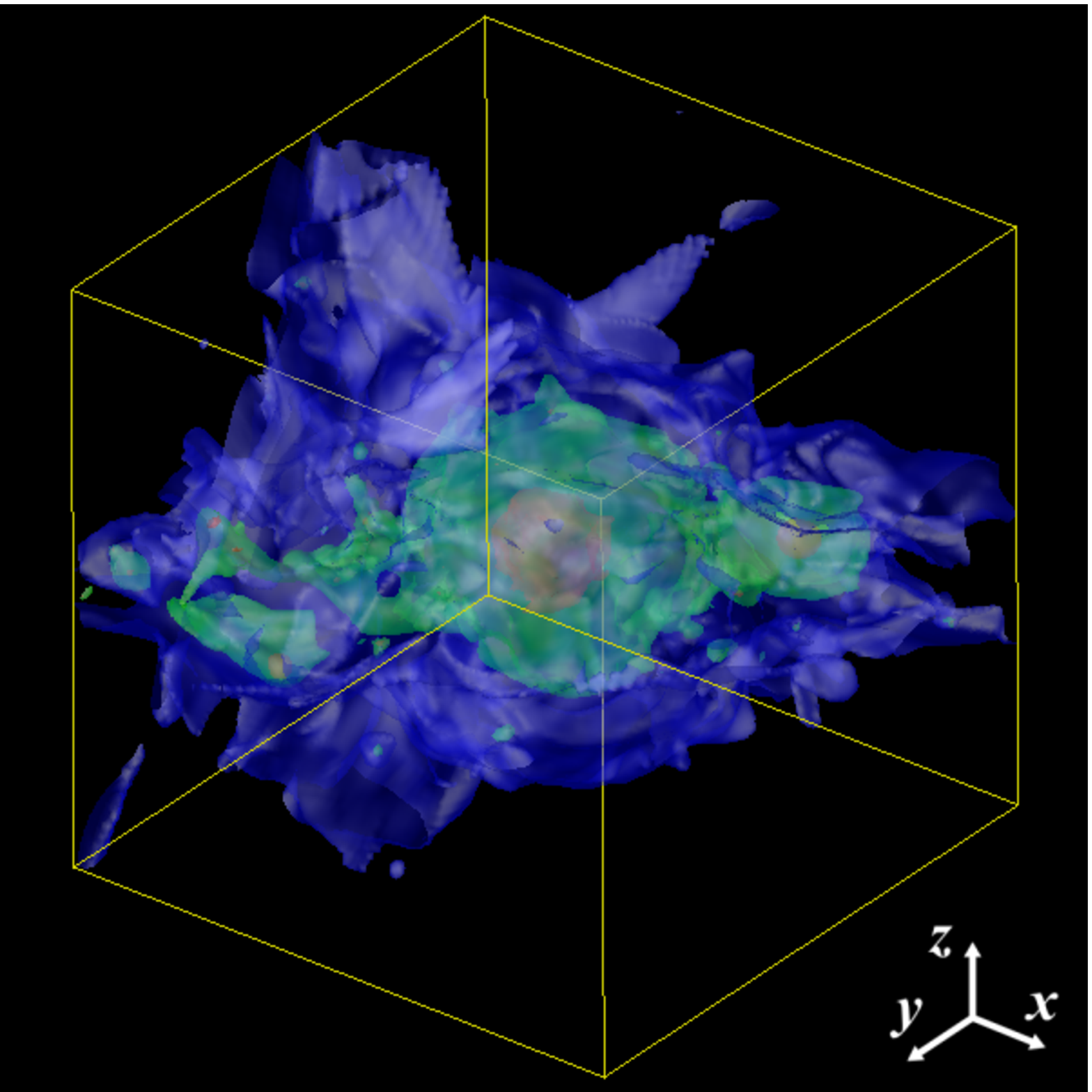}}}\\
\includegraphics[width=7.2cm]{fig1c.eps}\hspace{1cm}
\mbox{\raisebox{0cm}{\includegraphics[width=5.6cm]{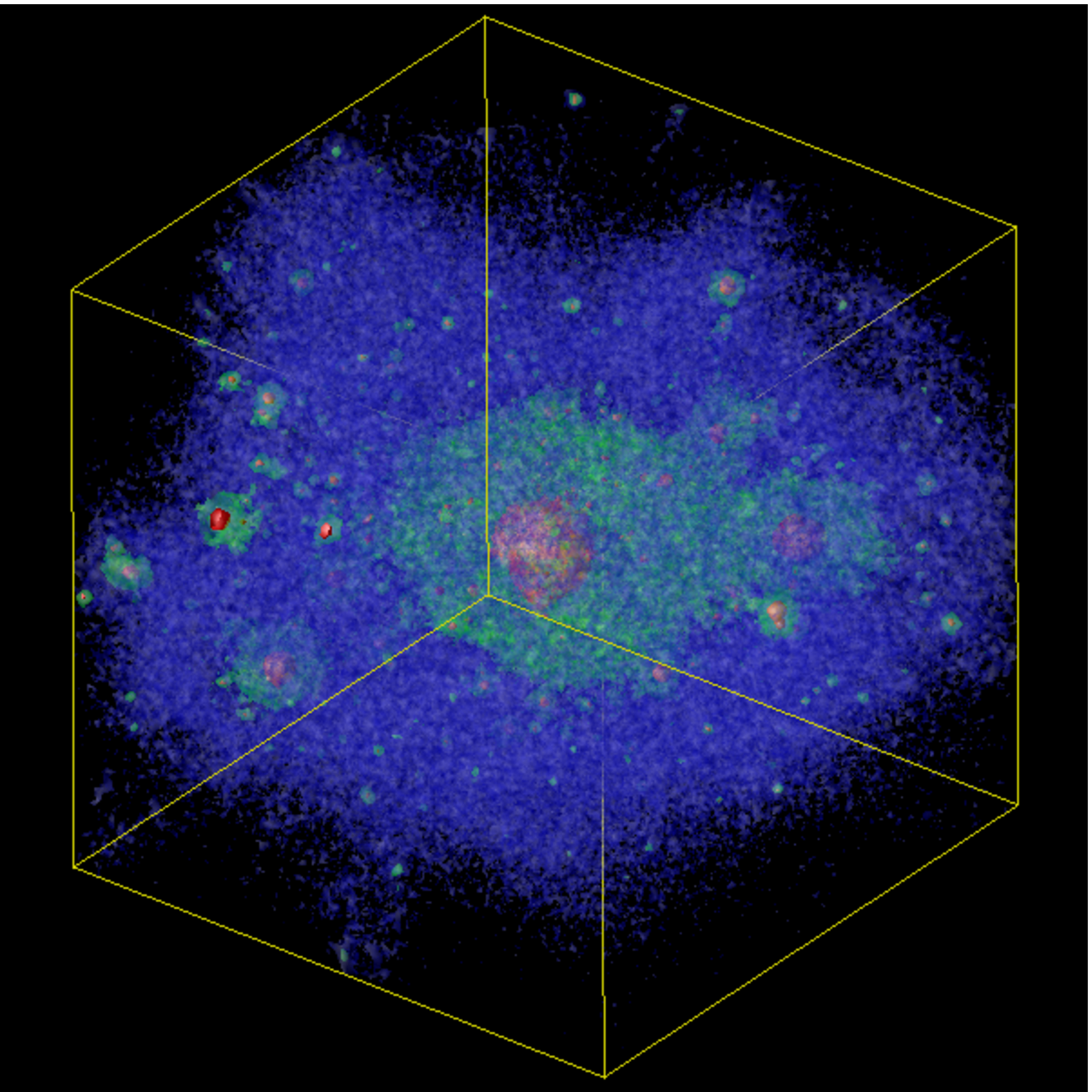}}}\\
\includegraphics[width=7.2cm]{fig1e.eps}\hspace{1cm}
\mbox{\raisebox{1.2cm}{\includegraphics[width=5.6cm]{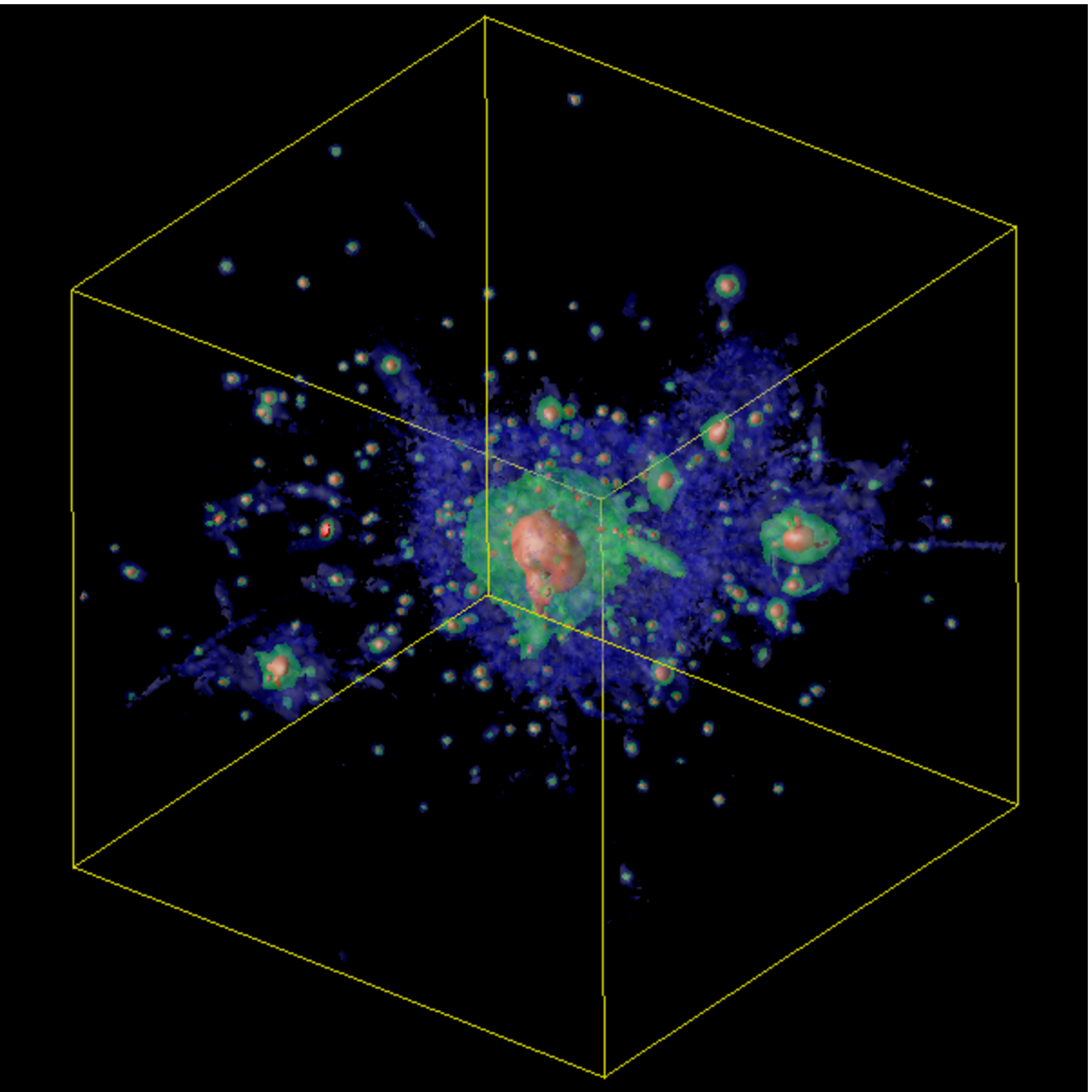}}}\\
\end{center}
\caption{Densities of gas (top), dark matter (middle) and stars (bottom) are shown in two ways for each. \textit{Left}: Projected surface densities on $x$-$z$ plane normalized by the fraction of each component in the box: $\log_{10}[\int dl\ \rho_k
\tilde{\Omega}_k^{-1}$ /(g cm$^{-3}$ $h^{-1}$ kpc)], where $k$ is one of gas, dark matter and stars and $\tilde{\Omega}_k$ is the fraction of $k$ component in the box.  \textit{Right}: Equal-density surfaces for $\log_{10}[\rho_k\tilde{\Omega}_k^{-1}$/(g cm$^{-3}$ $h^{-1}$ kpc)]=$-28.0$ (blue), $-27.0$ (green) and $-26.0$ (red). Densities are normalized in the same way as the left panel.}  \label{los}
\end{figure}

Projected surface densities of gas, dark matter and stars on $x$-$z$
plane are plotted in the left panels of Figure \ref{los}.  The right
panels of Figure \ref{los} show the three-dimensional view of the three
components. The left (right) plots are color-coded according to the
surface (space) densities normalized by the fraction of each component
averaged over the box, $\tilde{\Omega}_k$ ($k=$\, gas, dark matter and
stars). Note that the fraction $\tilde{\Omega}_k$ is different from the
density parameter $\Omega_k$ because the box is selected preferentially
around the cluster.

As is clear from Figure \ref{los}, the gas distribution is smoother but
very well traces the underlying dark matter distribution. In contract,
stars are more significantly concentrated in high density regions, and
exhibit numerous small clumps, most of which are not identified/resolved
in the gas distribution.

\begin{figure}
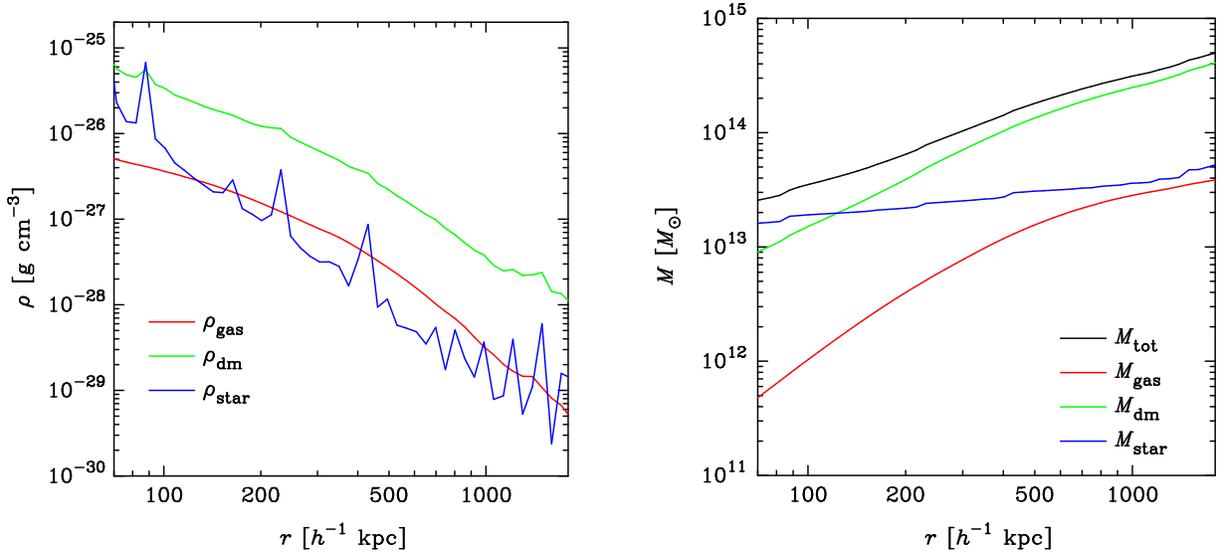

\begin{center}
\includegraphics[width=7.5cm]{fig2a.eps}\hspace{1cm}
\includegraphics[width=7.5cm]{fig2b.eps}
\end{center}
 \caption{Radial profiles of densities (left) and masses (right) are shown for gas (red), dark matter (green) and stars (blue). The black line in the right panel shows the total gravitational mass; $\Mtot=M_{\rm{gas}}+M_{\rm{dm}}+M_{\rm{star}}$. The analysis is performed on the 50 logarithmically equal radial bins.} \label{dens}
\end{figure}

Figure \ref{dens} plots the radial density and mass profiles of the
cluster. The stellar fraction in the inner region ($r<200 h^{-1}$kpc) is
significantly higher than the typical observed value. This is a
well-known common problem among current high-resolution cosmological
simulations, and implies that some important baryon physics including
high-energy phenomena and star formation is still missing in the
simulation.  We perform the analysis of cluster gas, {\it assuming} that
this excessive star densities in the inner region does not affect our
conclusions at outer radius.

\begin{figure}
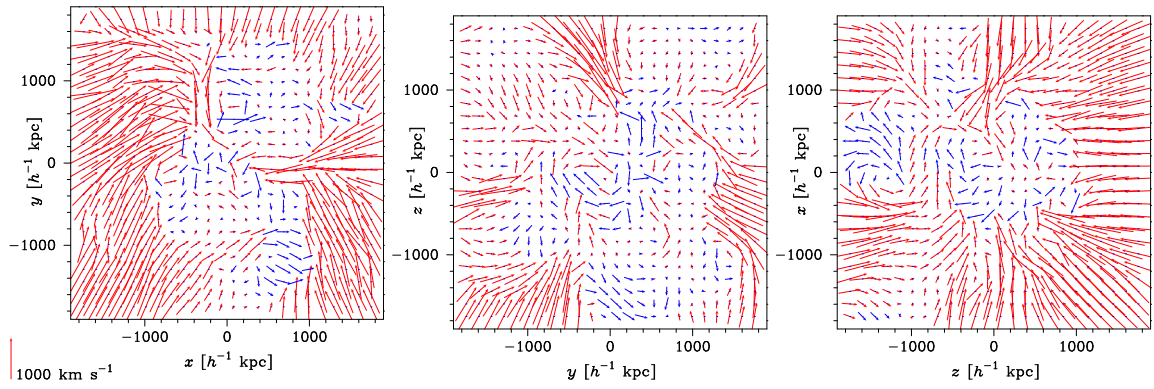

\begin{center}
\includegraphics[width=5cm]{fig3a.eps}
\includegraphics[width=5cm]{fig3b.eps}
\includegraphics[width=5cm]{fig3c.eps}
\end{center}
\caption{Velocity fields in $x$-$y$ (left), $y$-$z$ (middle) and $z$-$x$ (right) planes passing through the center of the cluster. The arrow is red if $v_r<0$, and blue if $v_r>0$. The length of the arrow is proportional to the magnitude of the velocity. An arrow with a speed of 1000 km s$^{-1}$ is shown for reference.}  \label{vel}
\end{figure}

Figure \ref{vel} represents velocity fields in $x$-$y$, $y$-$z$ and
$z$-$x$ planes passing through the center of the cluster. The red/blue
arrows have negative/positive radial velocity, showing that the gas in
the outer regions ($r\gtrsim 1 h^{-1}$ Mpc) falls toward the center
while its direction is randomized in the inner region.

\subsection{Results}

We evaluate the {\it effective} mass terms defined in Section 2 for the
simulated cluster, which is plotted in Figure \ref{gmass}; the left
panel shows the mass profiles, while the right panel indicates their
fractional contribution to the total mass within the radius.

The total mass $\Mtot(r)$ is computed by directly summing up all the
dark matter and star particles and gas of grids within the sphere of
$r$. The other terms, $\Mth$, $\Mrot$ and $\Mstr$, require the pressure
and velocity fields evaluated at $r$. For that purpose, we use the
density, velocity and temperature of gas defined at 520$^3$ original
grid points, and first bin the cluster into 50 logarithmically equal
radial intervals between 90--1900 $h^{-1}$ kpc, 10 $\times$ 10 linearly
equal angular intervals.

Integrands of equations (\ref{mthE}), (\ref{mrotE}) and (\ref{mstrE})
are calculated in each bin. Derivatives of the physical variables such
as $\del p/\del r$ are calculated as follows; first $\del/\del x$,
$\del/\del y$ and $\del/\del z$ are computed from the difference of the
adjacent the original Cartesian grid points. Then $\del/\del r$,
$\del/\del \theta$ and $\del/\del \varphi$ in our spherical coordinates
are calculated applying the chain rule.

\begin{figure}
\begin{center}
\includegraphics[width=7.5cm]{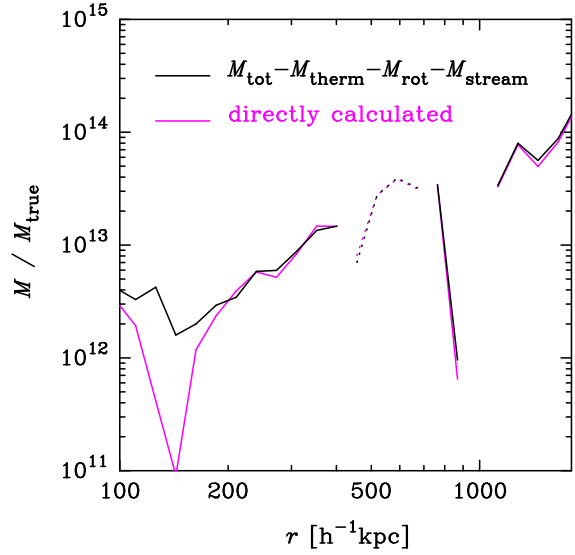}
\end{center}
 \caption{As comparison of the acceleration mass $\Macc$ calculated directly from the acceleration data (magenta) and $\Mtot-\Mth-\Mrot-\Mstr$ (black). Dotted line means that its sign is inverted. The analysis is performed on the 25 logarithmically equal radial bins.} \label{macc}
\end{figure}

In this way, $\Mth(r)$, $\Mrot(r)$ and $\Mstr(r)$ are computed by
integrating the corresponding integrands evaluated above.  Finally we
estimate $\Macc$ simply from the residual of
$\Macc=\Mtot-\Mth-\Mrot-\Mstr$, since we have the cluster data at $z=0$
alone. This estimate for $\Macc$ may be different from the original
defintion, i.e., equation (\ref{maccE}). Indeed when we attempted to
compute $\Macc$ directly from the box mentioned in Section 3, it turned
out to be too small to obtain a correct gravitational potential for the
entire cluster.  Thus we go back to a larger simulation box of a side of
22.5 $h^{-1}$ Mpc and the grid length of 29.34 $h^{-1}$ kpc that
encloses our cluster. Then we compute the gravitational potential using
FFT to obtain the gas acceleration at each grid point. This enables us
to directly calculate $\Macc$. Figure \ref{macc} is a comparison of
$\Macc$'s calculated by two methods. The directly calculated $\Macc$
(magenta line) is in good agreement with $\Mtot-\Mth-\Mrot-\Mstr$ (black
line), although there is a large difference between the two within $200
h^{-1}$ kpc. Also, we make sure that the sum $\Mth+\Mrot+\Mstr+\Macc$
reproduces $\Mtot$ within $\sim$ 2 \% except for the innermost region
($r<200 h^{-1}$ kpc), where it deviates from $\Mtot$ by up to $\sim$ 9
\%. Thus the estimation of $\Macc$ by $\Mtot-\Mth-\Mrot-\Mstr$ is
sufficiently good given the quoted errors of our conclusion
below. Although it seems better to use $\Macc$ directly calculated from
equation (\ref{maccE}), the grid size of the larger box is so coarse
that we cannot take advantage of the high resolution of the simulations
in this study. Therefore, we decided to use the smaller box explained in
Section 3 and $\Macc$ is calculated by $\Mtot-\Mth-\Mrot-\Mstr$ in the
following analysis.

The left panel of Figure \ref{gmass} implies that $\Mth$ agrees with
$\Mtot$ reasonably well. Each of the other three terms contributes less
than 10 \% of the total mass (the dotted curves correspond to the case
in which each term becomes negative and its absolute value is plotted
instead).

In order to consider the validity of HSE more quantitatively, we plot
the fractional contribution of each mass term in the right panel of
Figure \ref{gmass}. The rotation term, $\Mrot$, is always positive (by
definition) and contributes approximately 10 \% almost independently of
radius. In contrast, the streaming velocity term, $\Mstr$, is mostly
negative, and varies a lot at different radial bins. As a result, the
difference of the total mass $\Mtot$ and the HSE mass $\Mth$ is mostly
explained by the acceleration term $\Macc$ alone; compare the black and
magenta curves in the right panel of Figure \ref{gmass}.  At $r=r_{500}$
and $r_{200}$, the deviation from HSE in terms of the mass difference
$(\Mth-\Mtot)/\Mtot$ is about 10 \%. Nevertheless the value
significantly varies at different radii and it is safe to conclude that
$(\Mth-\Mtot)/\Mtot$ ranges approximately 10-20 \% at $r<r_{200}$. Also
there is no systematic trend of the validity of the HSE assumption as a
function of radius. Even though the reliability of the simulation is
suspicious for $r<200h^{-1}$kpc due to the excessive stellar
concentration (Section 3.1), $(\Mth-\Mtot)/\Mtot$ fluctuates between
$-10$ \% and $+25$ \% for $300h^{-1}{\rm kpc} < r < r_{500}$. Thus there
is no guarantee that HSE becomes a better approximation toward the inner
central region.

Physically speaking, $\Mrot+\Mstr+\Macc (=\Mtot-\Mth)$ corresponds to
the term intergrating the Lagrangian derivative of the gas velocity over
the sphere. Therefore the fact that it is small compared with $\Mth$ and
$\Mtot$ is simply translated into the condition of HSE that the gas
acceleration from a Lagrangian point of view is negligible compared with
the pressure gradient and the total gravity.

Since we analyze a single simulated cluster, it is not clear to what
extent our interpretation that $\Macc$ determines the departure from HSE
holds in general. Thus we use a different set of SPH simulation clusters
\citep{dolag09} kindly provided by Klaus Dolag. For these clusters, we
assume spherical symmetry and do not divide the spherical surface at a
given radius just for simplicity. Klaus Dolag also provides us with
$\Macc$ directly computed from the acceleration data.  We find that in
most cases $|\Macc|$ is larger than $\Mrot$ and $|\Mstr|$ especially
where $\Mth$ deviates from $\Mtot$. We also confirm that
$\Mth+\Mrot+\Mstr+\Macc$ reproduces $\Mtot$ within $\sim 5$ \% for most
regions (See Appendix C).

The above results basically support our conclusion for the cluster from
the AMR simulations, but we do not have a statistical discussion
including clusters from the SPH simulations because of the differences
in resolutions and simulation methods.  Instead, we divide the AMR
cluster into two regions; upper and lower hemispheres with respect to
the $x$-$y$ plane.  Then we duplicate each hemisphere into one
cluster. We call the synthetic cluster constructed from the $z>0$
($z<0$) hemisphere ``z+'' (``z-'').  Although these clusters are of
course not independent of the original cluster and we cannot make a
statistical argument on their properties, we can briefly look at the
effects of substructures or inhomogeneity of temperature and velocity
field.

We repeat the same analysis on these two synthetic clusters, and the
results are plotted in Figure \ref{zpm}. The amplitudes of the different
terms vary significantly between the two clusters, and the degree of the
validity of HSE is also very different. Nevertheless the generic trend
is clear; the relation of $\Macc \approx \Mtot-\Mth$ holds almost
independently of $r$.

It is not clear, however, why the two hemispheres have so different
values of $(\Mtot-\Mth)/\Mtot$; HSE holds very well for ``z+'', while it
is not the case for ``z-''. The visual inspection of Figure \ref{los}
does not reveal any significant difference between the two. It may be
because some local concentrations of dark matter enhance the
acceleration/deceleration of gas, and influence the overall
non-sphericity of the gas density. Thus the analysis taking account of
the ellipticity may provide a deeper insight on the validity of HSE, but
is beyond the scope of the present paper.

\section{Conclusion}

We have examined the validity of HSE that has been conventionally
assumed in estimating the mass of galaxy clusters from X-ray
observations. We use a simulated cluster and evaluate several mass terms
directly corresponding to the Euler equations that govern the gas
dynamics. We find that the mass estimated under the HSE assumption,
$\Mth$ in the present study, deviates from the true mass $\Mtot$ on
average by $\sim (10-20)$ \% fractionally for $r<r_{200}$. There is no
clear tendency that the HSE becomes a better approximation toward the
inner region. More importantly, we find that $\Mtot-\Mth$ is nearly
identical to $\Macc$, in other words, the validity of HSE is controlled
by the amount of gas acceleration. This trend is confirmed by the
separate analysis of the different hemispheres of the same simulated
cluster.

Our current analysis is limited to a single simulated cluster, but the
overall conclusion that the HSE mass agrees with the total mass within
(10-20)\% is consistent with previous results by \citet{fang09} and
\citet{lau09}. Nevertheless the interpretation of the origin of the
departure from HSE is very different.  \citet{fang09} concluded that the
gas rotation term $\Mrot$ makes a significant contribution and that
$\Mth+\Mrot$ well reproduces the total mass, especially for relaxed
clusters. It is not the case, however, for our simulated cluster at
least. Similarly \citet{lau09} found the similar degree of the departure
from HSE, but they ascribed the discrepancy to the random gas
motion. Their analysis, however, is based on the modification of the
Jeans equations, which does not appear to be justified for the analysis
of the gas dynamics, and thus their conclusion should be interpreted
with caution.

A relatively small systematic error of the HSE mass inferred from
current numerical simulations may be partly ascribed to the assumptions
inherent in the Euler equations, i.e., local thermal equilibrium and
negligible viscosity (Appendix A).  This is supported by the fact that
the error in the mass estimated from the random motion of collisionless
particles tends to be much greater at large radii (Appendix B) because
the relaxation time scale for collisionless particles are appreciably
longer than that for collisional gas. We should also note that the HSE
mass can be influenced by other physical precesses that are not included
in the numerical simulations, such as pressure support from
micro-turbulence, the magnetic field, and accelerated particles
\citep[e.g.,][]{lagana10}. The neglected components mentioned above are
closely linked with one another (e.g., viscosity can play a role in
generating turbulence and the magnetic field can affect both
thermalization and acceleration of gas particles) and will be
investigated in the near future by X-ray missions such as
NuSTAR\footnote{http://www.nustar.caltech.edu} and ASTRO-H as well as by
radio telescopes including EVLA\footnote{http://www.aoc.nrao.edu/evla}
and LOFAR\footnote{http://www.lofar.org}.

The present analysis is fairly idealized in a sense that the evaluation
of all the mass terms has been done from the full three dimensional data
of simulated clusters. In reality observational data of X-ray clusters
are basically projected along the line of sight. Therefore additional
uncertainties and systematic errors may become important as well. Also
it is necessary to carry out the current analysis for a number of
clusters in order to obtain the statistically robust conclusion.  These
issues will be discussed elsewhere.
 
\section*{ACKNOWLEDGMENTS}

We thank an anonymous referee for useful comments, 
in particular on the role of the unthermalized gas component as
discussed in Section 2.2. We also thank Klaus Dolag for kindly providing us
with SPH simulation data and calculating the gas acceleration, and
Fujihiro Hamba for a useful comment on the interpretation of the
inertial term in the Euler equations. Y.S. gratefully acknowledges
support from the Global Scholars Program of Princeton University. This
work is supported in part by the Grants-in-Aid for Scientific Research
by the Japan Society for the Promotion of Science (JSPS) (PD:22-5467,
21740139, 20340041, 24340035) and JSPS Core-to-Core Program
``International Research Network for Dark Energy''. Computing resources
were in part provided by the NASA High-End Computing (HEC) Program
through the NASA Advanced Supercomputing (NAS) Division at Ames Research
Center. The work of R.C. is supported in part by grants NNX11AI23G.


\appendix

\section{Relation between the Euler Equations and the Jeans Equations}

From a microscopic point of view, both the Euler equations and the Jeans equations can be derived from the Boltzmann equation under different assumptions. In the following, we explicitly compare the two sets of equations in both Cartesian and spherical coordinates. 

\subsection{Cartesian Coordinates}

We define the distribution function $f$ such that $f(\bs{x},\bs{v},t)d^3xd^3v$ is the probability that a randomly chosen particle in the system lies in the phase space volume $d^3xd^3v$ at position $(\bs{x},\bs{v})$ and time $t$.  The motion of such particles under the gravitational potential $\phi$ is described by the Boltzmann equation:
\begin{equation}
\pd{f}{t}+v^i\pd{f}{x^i}-\pd{\phi}{x_i}\pd{f}{v^i}
=\left(\frac{\delta f}{\delta t}\right)_{\rm{coll}}, 
\label{BE}
\end{equation}
where the collision term on the right hand side takes account of collisions between particles. Note that $v^i$ ($i=1, 2, 3$) represents a coordinate in the phase space and should not be confused with the velocity field at the spatial point $x^i$.  For simplicity, we assume that all particles have the same mass $m$ in the following.
 
First, we consider a collisionless case with $(\delta f /\delta t)_{\rm{coll}}=0$.  Multiplying equation (\ref{BE}) by $m$ and integrating it over the velocity space yield the continuity equation:
\begin{equation}
\pd{\rho}{t}+\pd{(\rho \bar{v}^i)}{x^i}=0,
\label{JCE}
\end{equation}
where
\begin{equation}
\rho(\bs{x},t)=\int d^3v\ mf(\bs{x},\bs{v},t)
\end{equation}
and we introduce the average over the velocity space:
\begin{equation}
\bar{q}(\bs{x},t)=\frac{1}{\rho(\bs{x},t)}\int d^3v\ mq(\bs{x}, \bs{v}, t)f(\bs{x},\bs{v},t)
\label{barred}
\end{equation}
for an arbitrary variable $q$ such as $v^i$. Multiplying equation (\ref{BE}) by $mv^j$ and integrating over the velocity space give the momentum equations:
\begin{equation}
\pd{(\rho \bar{v}^i)}{t}+\pd{\tau^{ij}_{\rm{J}}}{x^j}=-\rho\pd{\phi}{x_i},
\label{JME}
\end{equation}
where
\begin{equation}
\tau^{ij}_{\rm{J}}=\rho\overline{v^iv^j}=\rho\sigma^{2, ij}+\rho\bar{v}^i\bar{v}^j
\end{equation}
and $\sigma^{2, ij}$ is the velocity dispersion tensor. Equations (\ref{JCE}) and (\ref{JME}) reduce to the Jeans equations:
\begin{equation}
\pd{\bar{v}^i}{t}+\bar{v^j}\pd{\bar{v}^i}{x^j}=-\frac{1}{\rho}\pd{(\rho\sigma^{2, ij})}{x^j}-\pd{\phi}{x_i}
\end{equation}

Next, we consider a collisional case. Rigorous handling of the collisional term is rather complicated and simplified models are often used. A conventional one is the Bhartnagar-Gross-Krock (BGK) equation, which employs a linearized collisional term:
\begin{equation}
\pd{f}{t}+v^i\pd{f}{x^i}-\pd{\phi}{x_i}\pd{f}{v^i}=-\frac{f-f_0}{\tau},
\label{CBE}
\end{equation}
where $\tau$ is the relaxation time of the system considered, $f_0$ is the Maxwellian
distribution function characterized by the local temperature $T(\bs{x}, t)$:
\begin{equation}
f_0(\bs{x},\bs{v},t)=\left[\frac{m}{2\pi k_BT(\bs{x},t)}\right]^{3/2}\exp\left[-\frac{m(\bs{v}-\bar{\bs{v}}(\bs{x},t))^2}{2k_BT(\bs{x},t)}\right],
\end{equation}
and $k_B$ is the Boltzmann constant. If we assume that mean values of
conservatives such as mass and momentum are the same as those in local thermal equilibrium, the collision term vanishes in the continuity and momentum equations.  In local thermal equilibrium, pressure is defined from the diagonal components of $\sigma_{ij}^2$ by $p\delta_{ij} \equiv \rho\sigma_{ij}^2$ and the dispersion tensor can be written as
\begin{equation}
\tau^{ij}_{\rm{E}}= p\delta^{ij} + \rho \bar{v}^i\bar{v}^j. 
\end{equation}
If we replace $\tau_{\rm{J}}$ in equation (\ref{JME}) with $\tau_{\rm{E}}$ and combine them with equation (\ref{JCE}), we obtain the Euler equations:
\begin{equation}
\pd{\bar{v}^i}{t}+\bar{v}^j\pd{\bar{v}^i}{x^j}=-\frac{1}{\rho}\pd{p}{x_i}-\pd{\phi}{x_i}.
\end{equation}
The difference between the Euler and the Jeans equations resides only in the form of the dispersion tensor.

If we retain the off-diagonal components of the dispersion tensor, they can be interpreted as viscosity, and the equations reduce to the Navier-Stokes equations \citep{choudhuri98, chapman70}.

\subsection{Spherical coordinates}
One can rewrite the continuity and momentum equations in the previous section in general coordinates:
\begin{equation}
\dt{\rho}+\nabla_i(\rho \bar{v}^i)=0
\end{equation}
and
\begin{equation}
\dt{(\rho\bar{v}^i)}+\nabla_j\tau^{ij}=-\rho g^{ij}\nabla_j\phi, 
\end{equation}
by using the \textit{covariant} derivative operator $\nabla_i$ . For spherical
coordinates ($x^1=r, x^2=\theta, x^3=\varphi$), non-zero components of
the metric tensor $g^{ij}$ are
\begin{equation}
g_{11}=1,\gap g_{22}=r^2,\gap g_{33}=r^2\sin^2\theta
\end{equation}
and the corresponding non-zero connection coefficients are
\begin{equation}
\begin{split}
&\Gamma^1_{22}=-r,\gap\Gamma^1_{33}=-r\sin^2\theta,\gap\Gamma^2_{12}=\Gamma^2_{21}=\frac{1}{r},\\ &\Gamma^2_{33}=-\sin\theta\cos\theta,\gap\Gamma^3_{13}=\Gamma^3_{31}=\frac{1}{r},\gap\Gamma^3_{23}=\Gamma^3_{32}=\cot\theta.
\end{split}
\end{equation}
The velocity vector is now given by
$\bar{v}^i=(\bar{v}_r, \bar{v}_\theta/r, \bar{v}_\varphi/r\sin\theta)$.

In spherical coordinates, the continuity equation reads 
\begin{equation}
\dt{\rho}+\frac{1}{r^2}\dr{(r^2\rho\bar{v}_r)}+\frac{1}{r\sin\theta}\dth{(\sin\theta\rho\bar{v}_\theta)}+\frac{1}{r\sin\theta}\dph{(\rho\bar{v}_\varphi)}=0.
\end{equation}
Setting $\tau^{ij}=\tau^{ij}_{\rm{E}}=\rho\bar{v}^i\bar{v}^j+pg^{ij}$ gives the 
Euler equations:
\begin{equation}
\left[\dt{}+\bar{v}_r\dr{}+\frac{\bar{v}_\theta}{r}\dth{}+\frac{\bar{v}_\varphi}{r\sin\theta}\dph{}\right]\bar{v}_r-\frac{\bar{v}_\theta^2+\bar{v}_\varphi^2}{r}=-\frac{1}{\rho}\dr{p}-\dr{\phi}
\label{reuler}
\end{equation}
\begin{equation}
\left[\dt{}+\bar{v}_r\dr{}+\frac{\bar{v}_\theta}{r}\dth{}+\frac{\bar{v}_\varphi}{r\sin\theta}\dph{}\right]\bar{v}_\theta+\frac{\bar{v}_r\bar{v}_\theta-\bar{v}_\varphi^2\cot\theta}{r}=-\frac{1}{\rho r}\dth{p}-\frac{1}{r}\dth{\phi}
\end{equation}
\begin{equation}
\left[\dt{}+\bar{v}_r\dr{}+\frac{\bar{v}_\theta}{r}\dth{}+\frac{\bar{v}_\varphi}{r\sin\theta}\dph{}\right]\bar{v}_\varphi+\frac{\bar{v}_r\bar{v}_\varphi+\bar{v}_\theta \bar{v}_\varphi\cot\theta}{r}=-\frac{1}{\rho r\sin\theta}\dph{p}-\frac{1}{r\sin\theta}\dph{\phi}.
\end{equation}
On the other hand, putting
$\tau^{ij}=\tau^{ij}_{\rm{J}}=\rho\overline{v^iv^j}$ leads to the Jeans
equations
\begin{equation}
\begin{split}
&\left[\dt{}+\bar{v}_r\dr{}+\frac{\bar{v}_\theta}{r}\dth{}+\frac{\bar{v}_\varphi}{r\sin\theta}\dph{}\right]\bar{v}_r+\frac{1}{\rho}\left[\dr{(\rho\sigmarr^2)}+\frac{1}{r}\dth{(\rho\sigmart^2)}+\frac{1}{r\sin\theta}\dph{(\rho\sigmarp^2)}\right]\\
&+\frac{1}{r}\left(2\sigmarr^2-\sigmatt^2-\sigmapp^2-\bar{v}_\theta^2-\bar{v}_\varphi^2+\sigmart^2\cot\theta\right)
=-\dr{\phi}
\end{split}
\label{jeansR}
\end{equation}
\begin{equation}
\begin{split}
&\left[\dt{}+\bar{v}_r\dr{}+\frac{\bar{v}_\theta}{r}\dth{}+\frac{\bar{v}_\varphi}{r\sin\theta}\dph{}\right]\bar{v}_\theta+\frac{1}{\rho}\left[\dr{(\rho\sigmart^2)}+\frac{1}{r}\dth{(\rho\sigmatt^2)}+
\frac{1}{r\sin\theta}\dph{(\rho\sigmatp^2)}\right]\\
&+\frac{1}{r}\left(3\sigmart^2-\sigmapp^2\cot\theta+\bar{v}_r\bar{v}_\theta-\bar{v}_\varphi^2\cot\theta+\sigmatt^2\cot\theta\right)
=-\frac{1}{r}\dth{\phi}
\end{split}
\end{equation}
\begin{equation}
\begin{split}
&\left[\dt{}+\bar{v}_r\dr{}+\frac{\bar{v}_\theta}{r}\dth{}+\frac{\bar{v}_\varphi}{r\sin\theta}\dph{}\right]\bar{v}_\varphi+\frac{1}{\rho}\left[\dr{(\rho\sigmarp^2)}+\frac{1}{r}\dth{(\rho\sigmatp^2)}+\frac{1}{r\sin\theta}\dph{(\rho\sigmapp^2)}\right]\\
&+\frac{1}{r}\left(3\sigmarp^2+\bar{v}_r\bar{v}_\varphi+\bar{v}_\theta\bar{v}_\varphi\cot\theta+2\sigmatp^2\cot\theta\right)
=-\frac{1}{r\sin\theta}\dph{\phi}.
\end{split}
\end{equation}


\section{Systematic Errors in Mass Estimates for Collisionless Systems}

In a similar fashion to Section 2.1, we can compute the gravitational mass using the Jeans equations:
\begin{equation}
\dt{\bs{v}}+(\bs{v}\cdot\nabla)\bs{v}=-\frac{1}{\rhod}{\nabla(\rhod\bs{\sigma}^2)}-\nabla{\phi},
\end{equation}
\begin{equation}
\Mtot=\frac{1}{4\pi G}\int_{\del V} d\bs{S}\cdot\left[-\frac{1}{\rhod}\nabla (\rhod\bs{\sigma}^2)-(\bs{v}\cdot\nabla)\bs{v}-\dt{\bs{v}}\right],
\label{massjeans}
\end{equation}
where $\bs{v}$ and $\bs{\sigma}^2$ are the velocity field of particles and the velocity dispersion tensor, respectively. We here represent the collisionless component by dark matter, but the same formulation is readily applicable to galaxies. 
We decompose the right hand side of equation (\ref{massjeans}) into the following terms by means of equation (\ref{jeansR}):
\begin{equation}
\Mtot=\Mrand+\Maniso+\Mrot+\Mstr+\Mcross+\Macc.
\label{jeansdecomp}
\end{equation}
\begin{equation}
\Mrand=-\frac{1}{4\pi G}\int_{\del V} dS\ \frac{1}{\rhod}\dr{(\rhod\sigmarr^2)},
\end{equation}
\begin{equation}
\Maniso=-\frac{1}{4\pi G}\int_{\del V} dS\ \frac{2\sigmarr^2-\sigmatt^2-\sigmapp^2}{r},
\label{maniso}
\end{equation}
\begin{equation}
\Mrot=\frac{1}{4\pi G}\int_{\del V} dS\ \frac{v_\theta^2+v_\varphi^2}{r},
\end{equation}
\begin{equation}
\Mstr=-\frac{1}{4\pi G}\int_{\del V} dS\ \left[v_r\dr{v_r}+\frac{v_\theta}{r}\dth{v_r}+\frac{v_\varphi}{r\sin\theta}\dph{v_r}\right],
\end{equation}
\begin{equation}
\Mcross=-\frac{1}{4\pi G}\int_{\del V} dS\ \left[\frac{1}{\rhod r}\dth{(\rhod\sigmart^2)}+\frac{1}{\rhod r\sin\theta}\dph{(\rhod\sigmarp^2)}+\frac{\sigmart^2\cot\theta}{r}\right].
\end{equation}
\begin{equation}
\Macc=-\frac{1}{4\pi G}\int_{\del V} dS\ \dt{v_r}.
\end{equation}
Physical interpretation of each mass term is as follows. The term $\Mrand$ comes from the gradient of velocity dispersion in the $r$-direction and corresponds to $\Mth$ for a collisional gas. The meaning of $\Mrot$, $\Mstr$ and $\Macc$ are similar to the corresponding terms for the collisional gas (equations (6) to (8)). The terms that have no counterpart in the Euler equations are $\Maniso$ and $\Mcross$; the former represents anisotropy of the velocity dispersion whereas the latter arises from the off-diagonal components of the velocity dispersion tenser and vanishes if velocities of different directions are uncorrelated.

We apply the above formulation to dark matter particles in the simulated cluster described in Section 3 to quantify intrinsic systematic errors of the mass estimation using collisionless particles. Note that one can apply the same method to galaxies but with much larger impact of statistical errors. Therefore, we do not do so here because we are interested in intrinsic systematic errors independent of observational complexities. Each term is computed in a similar manner to the case of collisional gas described in Section 3. 

\begin{figure}
\begin{center}
\includegraphics[width=7.5cm]{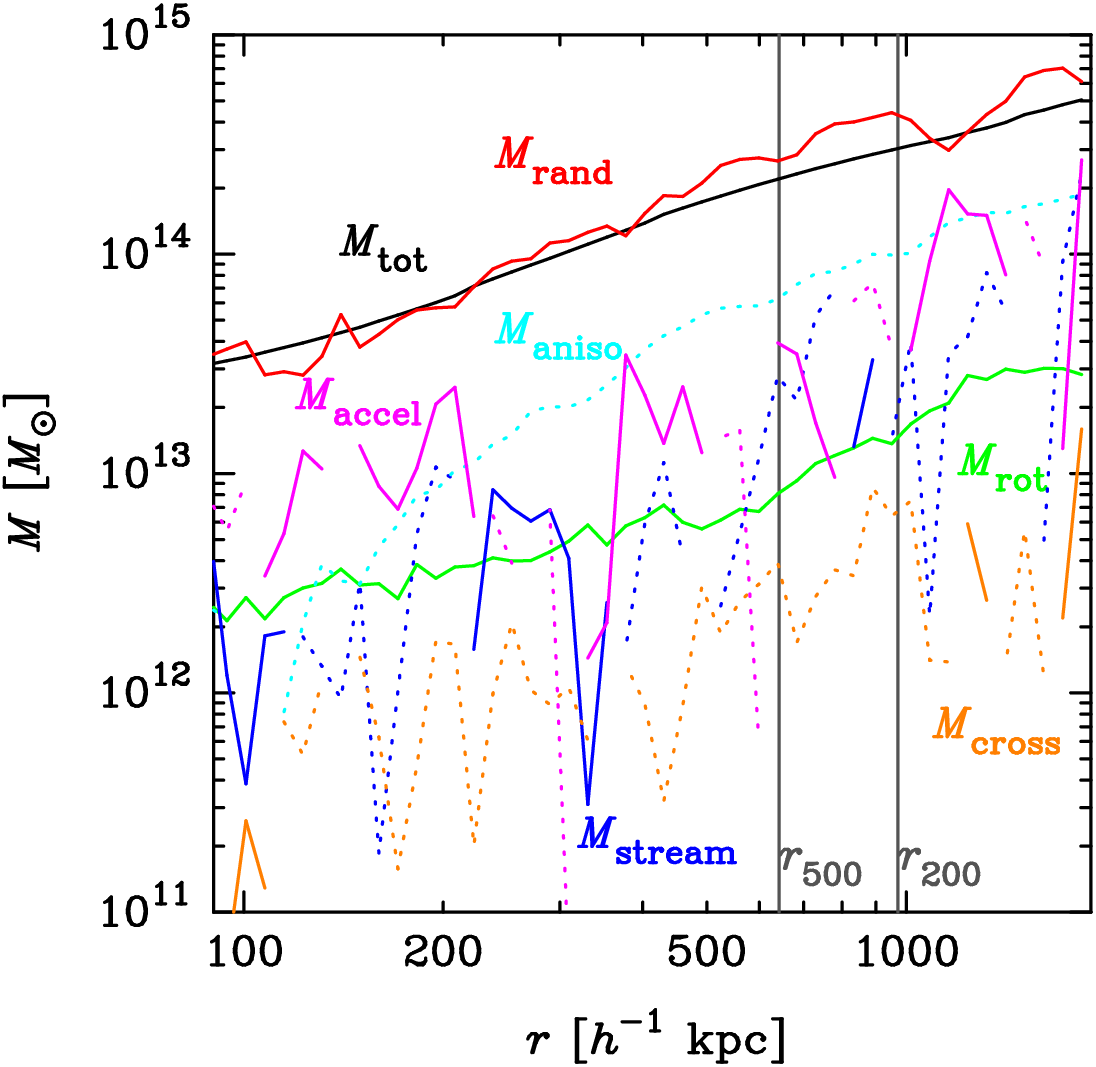}\hspace{1cm}
\includegraphics[width=7.5cm]{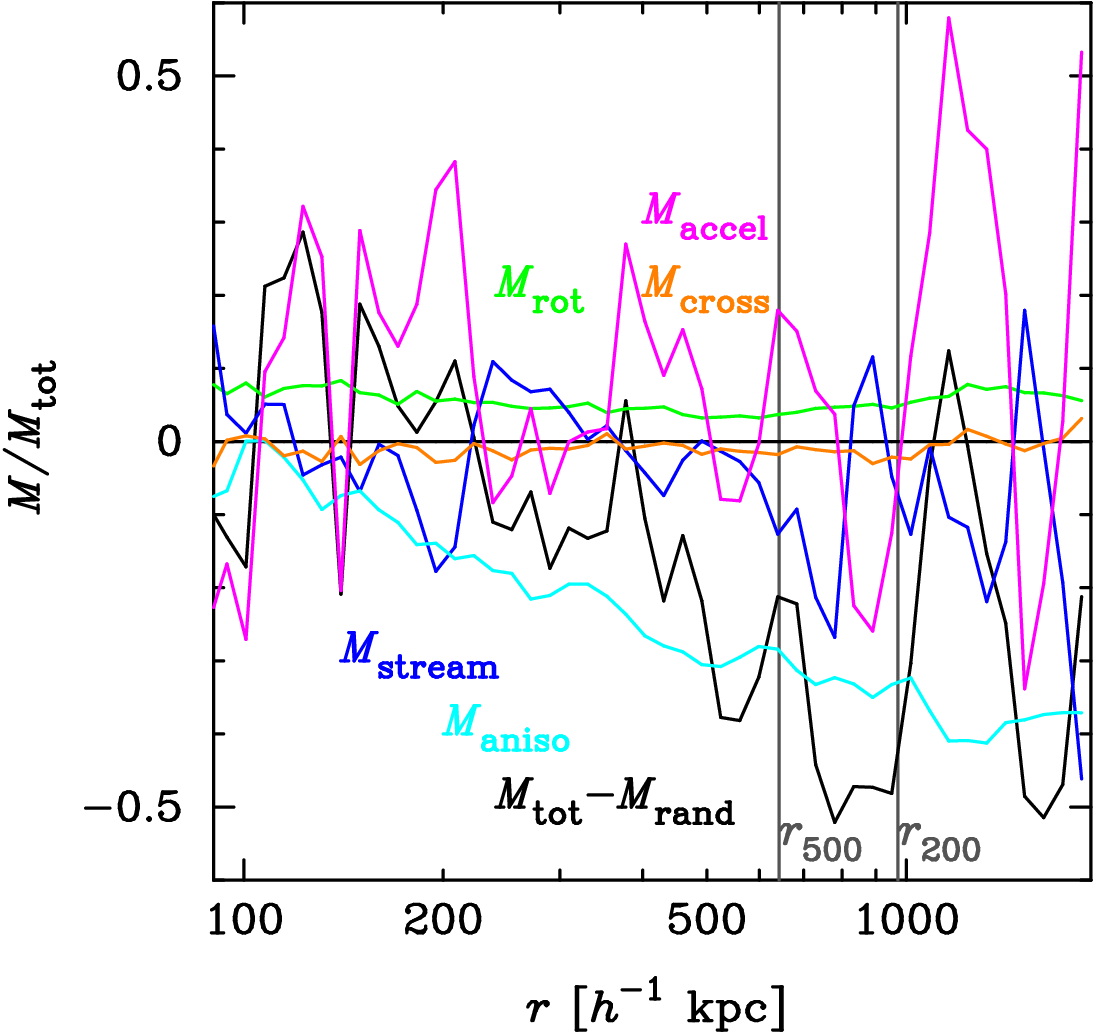}
\end{center}
\caption{The \textit{effective} mass terms in Equation (\ref{jeansdecomp}) for the dark matter in the simulated cluster are shown in the left panel: $\Mtot$ (black), $\Mrand$ (red), $\Mrot$ (green), $\Mstr$ (blue), $\Maniso$ (cyan), $\Mcross$ (orange) and $\Macc$ (magenta). Here $\Macc$ is calculated by $\Macc=\Mtot-\Mrand-\Mrot-\Mstr-\Maniso-\Mcross$. Dotted line means that its sign is inverted. Ratios of mass terms to $\Mtot$ are shown in the right panel;  The black line shows $(\Mtot-\Mrand)/\Mtot$ and other mass terms are colored in the same colors as the left panel.}
\label{dmass}
\end{figure}

Figure \ref{dmass} shows that the difference between $\Mtot$ and $\Mrand$ increases toward the outer envelope mainly owing to the presence of $\Maniso$.  This is because the relaxation timescale of collisionless particles is much longer than that of the collisional gas. Once this term is subtracted, $\Mtot - \Mrand - \Maniso$ closely matches $\Macc$ whose absolute value is limited to within $\sim 0.3 \Mtot$.  The amount of $\Macc$ is similar to that for the collisional
gas (Fig. \ref{gmass}). The other mass terms such as $\Mcross$ are less important.  

The above results imply that proper account of velocity anisotropies is essential for the mass reconstruction using a collisionless component. We stress that $\Maniso$ is irrelevant to the collisional fluid as long as local thermal equilibrium is established (Appendix A).

\section{Validity of Hydrostatic Equilibrium in Clusters from Smoothed Particle Hydrodynamics Simulation}

We have shown the behavior of the mass terms for the cluster from the AMR simulation in Section 4. Especially, the HSE mass deviates from the total mass by up to $\sim$ 30 \% and the difference is explained chiefly by the gas acceleration. In order to make sure if such behavior is common to simulated clusters, we analyze the other simulated clusters. Since our AMR simulation includes only a single cluster, we analyze clusters extracted from a smoothed particle hydrodynamics (SPH) simulation performed by \citet{dolag09} using GADGET-2 to confirm the results in Section 4. We refer \citet{dolag09} for the detail of the simulation.

We use five regions named g1, g72, g1542, g3344 and g914 extracted from the simulation. Each region has a cluster near the center and the cluster is labeled ``a''  (g1a, g72a, g1542a, g3344a, g914a). For each cluster, we make $256^3$ mesh data centered on the cluster's center-of-mass within the radius $r_{500}$. The size of the data box is determined so that it contains the entire sphere of radius $3 r_{500}$ centered on the cluster's center-of-mass. 

We calculate the mass terms defined in Section 2 for the five simulated clusters. Since we have the gas acceleration data, we calculate $\Macc$ directly from the data without using $\Mtot-\Mth-\Mrot-\Mstr$. 

\begin{figure}
\begin{center}
\includegraphics[width=7.5cm]{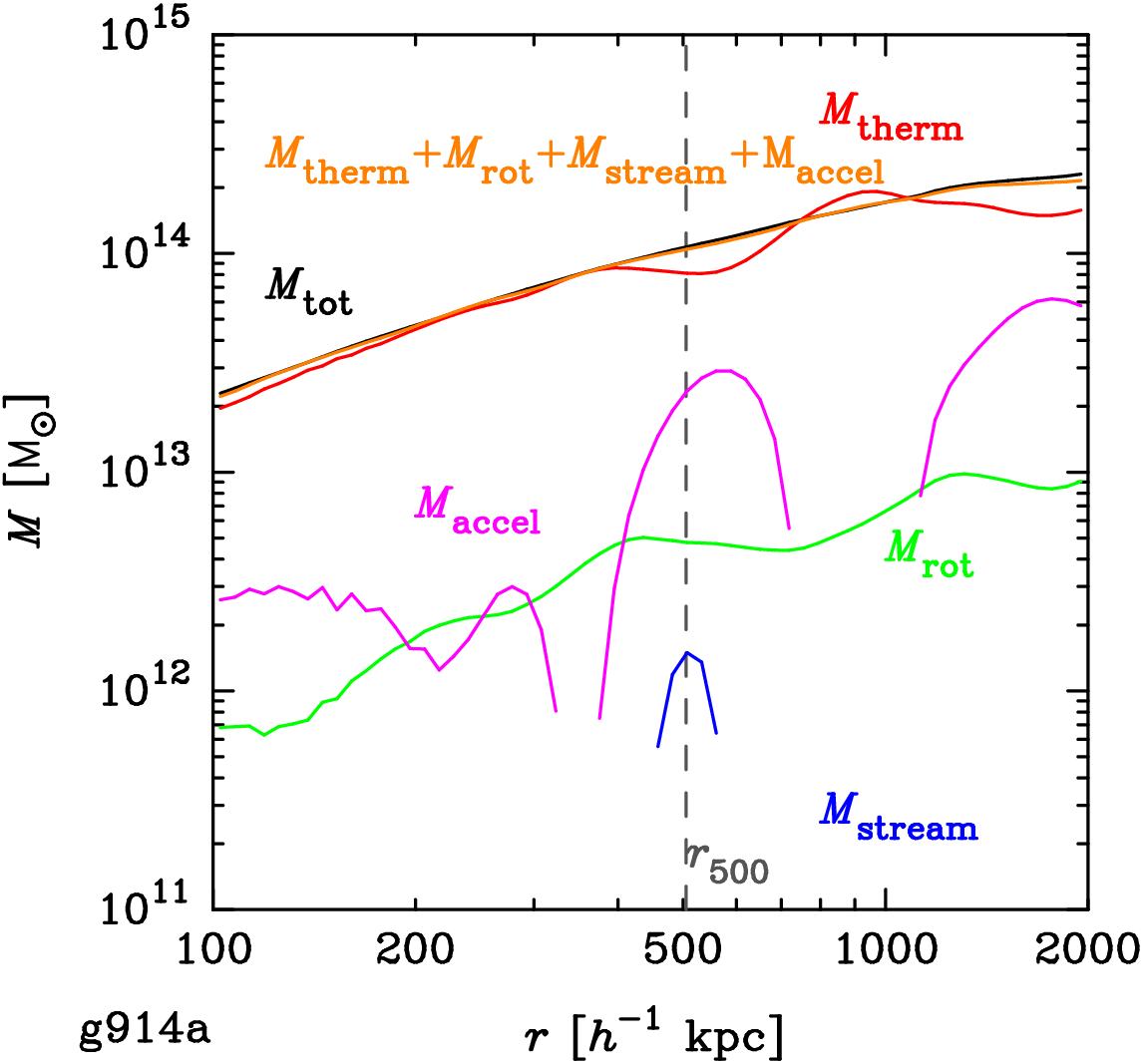}\hspace{1cm}
\includegraphics[width=7.5cm]{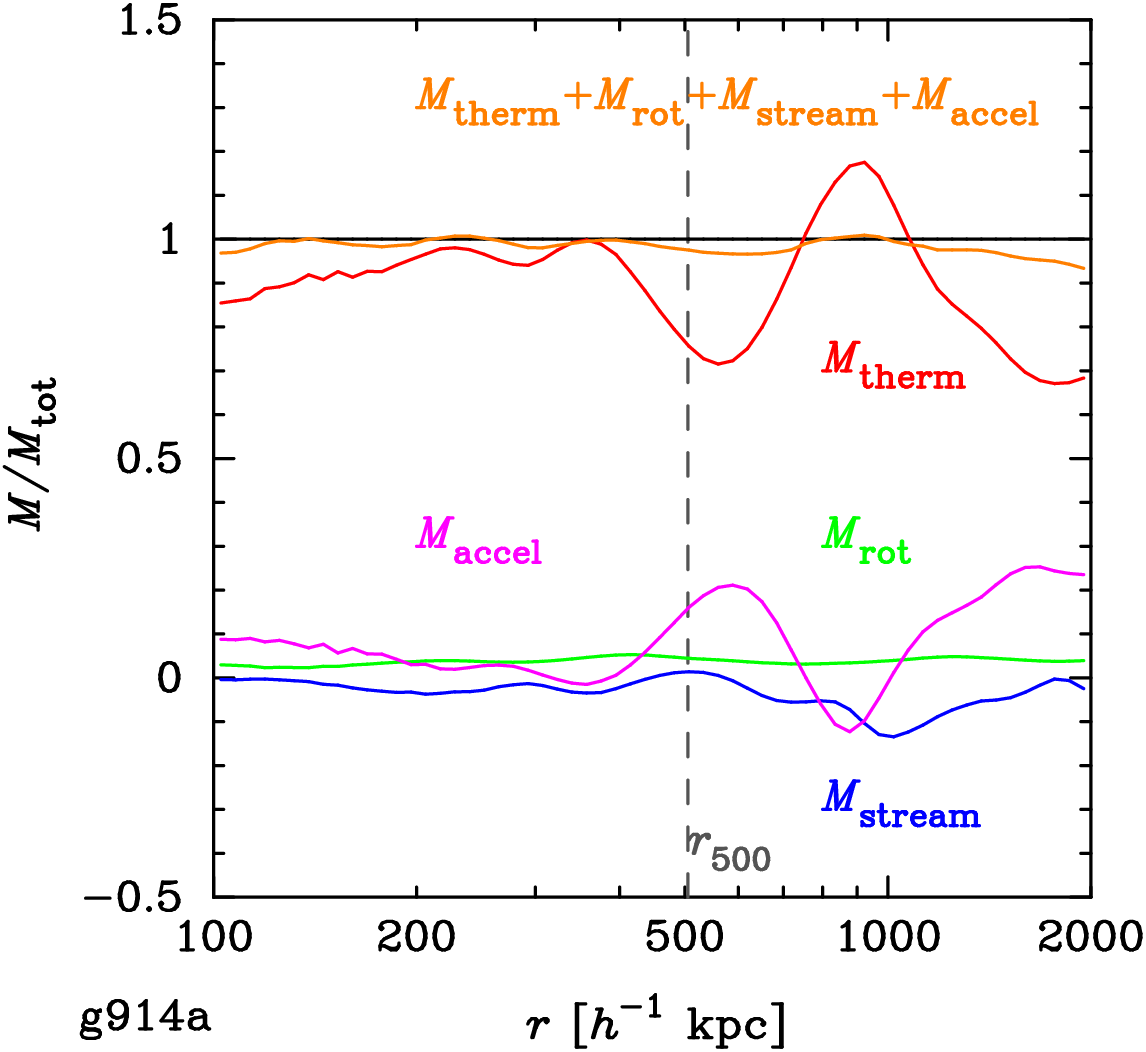}
\end{center}
\caption{The \textit{effective} mass terms in Equation (\ref{eulerdecomp}) for the cluster g914a from the SPH simulation are shown in the left panel: $\Mtot$ (black), $\Mth$ (red), $\Mrot$ (green), $\Mstr$ (blue), $\Maniso$ (cyan), $\Mcross$ (orange) and $\Macc$ (magenta). It also shows the sum  $\Mth+\Mrot+\Mstr+\Macc$ in orange. Here $\Macc$ is calculated using the acceleration data, not $\Macc=\Mtot-\Mth-\Mrot-\Mstr$. Ratios of mass terms to $\Mtot$ are shown in the right panel. The color-coding is the same as the left panel. The analysis is performed on the 60 logarithmically equal radial bins.}
\label{sph}
\end{figure}

Figure \ref{sph} shows the result for the cluster g914a. This figure basically supports the results in Section 4 in that $\Mth$ deviates from $\Mtot$ by $\sim 30\%$ at most and that $\Macc$ becomes large where the difference between $\Mtot$ and $\Mth$ is large. The fact that the sum $\Mth+\Mrot+\Mstr+\Macc$ is approximately $\Mtot$ means the estimation of $\Macc$ by $\Mtot-\Mth-\Mrot-\Mstr$ is good, as confirmed in Section 4 for the AMR cluster.

Although not graphically shown, the mass terms for the other simulated clusters also exhibit similar behavior. 

\begin{figure}
\begin{center}
\includegraphics[width=7.5cm]{fig5a.eps}\hspace{1cm}
\includegraphics[width=7.5cm]{fig5b.eps}
\end{center}
 \caption{The \textit{effective} mass terms in Equation (\ref{eulerdecomp}) for the gas in the simulated cluster are shown in the left panel: $\Mtot$ (black), $\Mth$ (red), $\Mrot$ (green), $\Mstr$ (blue) and $\Macc$ (magenta). Here $\Macc$ is calculated by $\Macc=\Mtot-\Mth-\Mrot-\Mstr$. Dotted line means that its sign is inverted. Ratios of mass terms to $\Mtot$ are shown in the right panel. The black line shows $(\Mtot-\Mth)/\Mtot$ and colored lines represent the same things as the left panel. The analysis is performed on the 50 logarithmically equal radial bins and 10 linearly equal bins both in polar and azimuthal angles.} \label{gmass}
\end{figure}

\begin{figure}
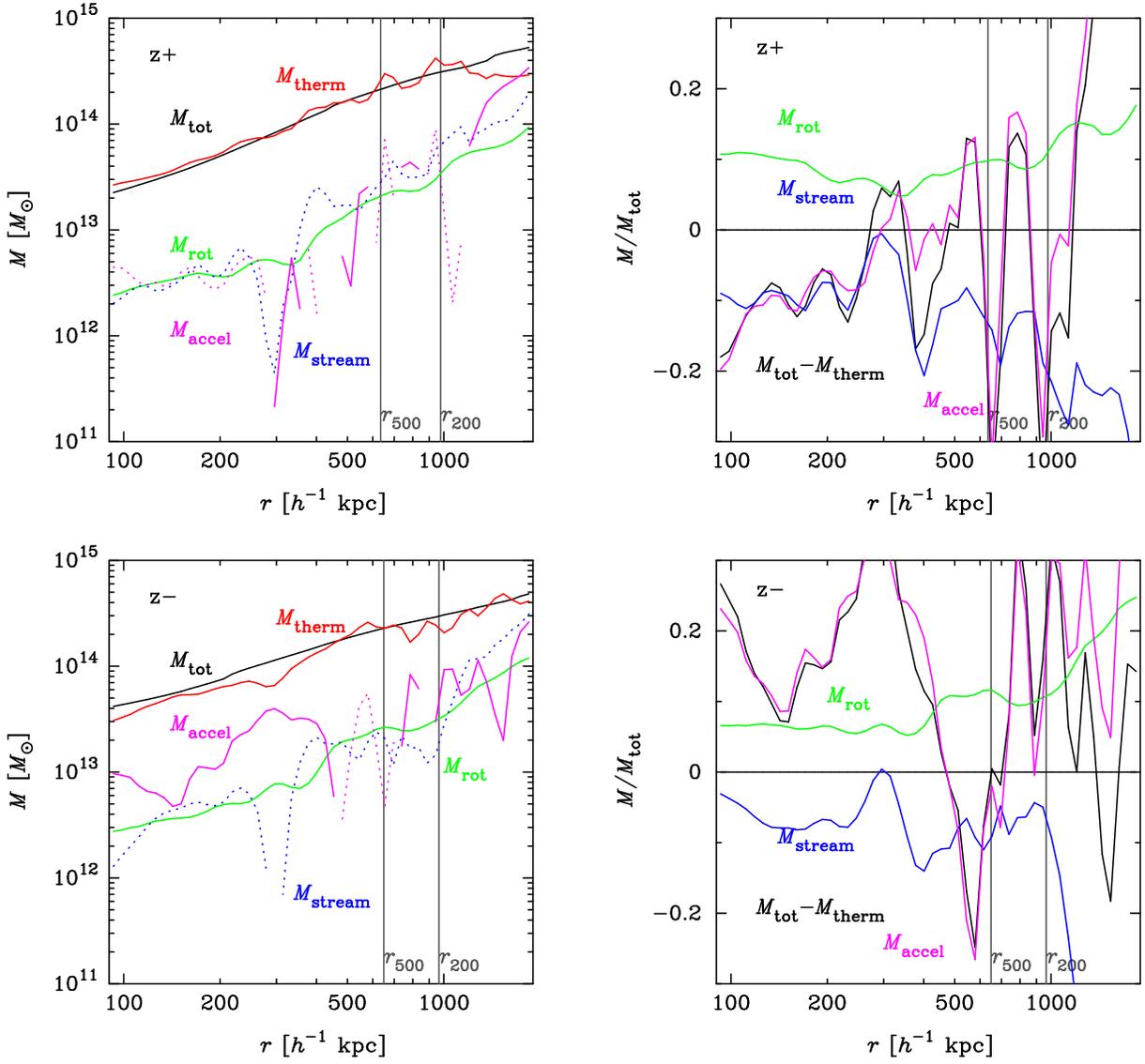

\begin{center}
\includegraphics[width=7.5cm]{fig6a.eps}\hspace{1cm}
\includegraphics[width=7.5cm]{fig6b.eps}
\end{center}
\begin{center}
\includegraphics[width=7.5cm]{fig6c.eps}\hspace{1cm}
\includegraphics[width=7.5cm]{fig6d.eps}
\end{center}
\caption{Same as Figure \ref{gmass}, but for the clusters ``z+'' (top) and ``z-'' (bottom).} \label{zpm}
\end{figure}

\end{document}